\documentclass[aps,amsmath,amssymb,twocolumn,superscriptaddress]{revtex4-1}

\usepackage{graphicx}
\usepackage{bm}
\usepackage{dsfont}
\usepackage{array}
\usepackage[usenames,dvipsnames]{xcolor}
\usepackage{setspace}
\usepackage[compact]{titlesec}

\usepackage{epstopdf}

\begin{document}


\title{Electrophoresis of Janus Particles: a Molecular Dynamics simulation study}

\author{Taras Y. Molotilin}
\affiliation{A.N. Frumkin Institute of Physical Chemistry and Electrochemistry, Russian Academy of Sciences, 31 Leninsky Prospect, 119071 Moscow, Russia}

\author{Vladimir Lobaskin}
\affiliation{School of Physics and Complex and Adaptive Systems Lab, University College Dublin, Belfield, Dublin 4, Ireland}

\author{Olga I. Vinogradova}
 \email{oivinograd@yahoo.com}
\affiliation{A.N. Frumkin Institute of Physical Chemistry and Electrochemistry, Russian Academy of Sciences, 31 Leninsky Prospect, 119071 Moscow, Russia}
\affiliation{Department of Physics, M. V. Lomonosov Moscow State University, 119991 Moscow, Russia}
\affiliation{DWI - Leibniz Institute for Interactive Materials, RWTH Aachen, Forckenbeckstra\ss e 50, 52056 Aachen, Germany}

\date{\today}

\begin{abstract}
In this work, we use Molecular Dynamics and Lattice-Boltzmann simulations to study the properties of charged Janus particles in electric field. We show that for relatively small net charge and thick electrostatic diffuse layer mobilities of Janus particles and uniformly charged colloids of the same net charge are identical. However, for higher charges and thinner diffuse layers Janus particles always show lower electrophoretic mobility. We also demonstrate that Janus particles align with the electric field and the angular deviation from the field's direction is related to their dipole moment. We show that the latter is affected by the thickness of electrostatic diffuse layer and strongly correlates with the electrophoretic mobility.
\end{abstract}

\maketitle

\section{\label{sec:level1}Introduction}

Electrophoresis is both a useful tool and a broad field of research that has recently met its 200th anniversary.\cite{hunter81,lyklema95} Since then, much work has been done, and today numerous applications exist, and often are even treated as somewhat routine.

Until recently, most studies of electrophoresis have assumed that particles are uniformly charged. In such a situation, the electrophoretic  mobility $\mu$, which relates the  translational velocity $v_c$ of a particle of radius $R$ immersed in electrolyte solution of concentration $C_{\Sigma}$ to the electric field $\mathbf{v_c} = \mu \mathbf{E}$, is given by $\mu\left(\zeta\right)$. Here, $\zeta$ is the zeta potential, which for hydrophilic surfaces is simply equal to surface electrostatic potential determined by the charge density of the particle (but note that for hydrophobic particles the situation is more complicated).\cite{Maduar2015} The exact $\mu\left(\zeta\right)$ relation depends on the thickness of the electrostatic diffuse layer (EDL) via a dimensionless quantity $\kappa R$, where $\kappa$ is the inverse Debye length, $\kappa=\left(4 \pi l_B C_{\Sigma}\right)^{1/2}$ with $l_B$  being the Bjerrum length and $C_{\Sigma}$ for a 1:1 electrolyte is the total concentration of ions in the system. The dimensionless mobility, $\widetilde{\mu}=6\pi\eta l_B\mu/e$, where $\eta$ stands for the dynamic viscosity of the solvent and $e$ is the elementary charge, can be expressed as
\begin{equation}
	\label{eq:mu2}
	\widetilde{\mu} = f \widetilde{\zeta},
\end{equation}
with dimensionless zeta potential $\widetilde{\zeta}=\zeta e/k_BT$ (where $k_BT$ denotes the thermal energy), which can be deduced from measured $\widetilde{\mu}$ if $f$ is known. Earlier models have predicted $f=1$ in the H\"{u}ckel thick EDL limit\cite{Huckel1924} ($\kappa R \ll 1$) and $f=3/2$ in the Smoluchowski thin EDL limit\cite{Smoluchowski1924} ($\kappa R \gg 1$).
To calculate $f$ depending on $\kappa R$ the majority of previous works have used the classical mean-field solution by O'Brien and White,\cite{OBrienWhite1978} which is often referred to as the \emph{Standard Electrokinetic Model} (SEM). This theory has played a major role in the interpretation of electrophoretic measurements over several decades.

The assumption that particles are uniformly charged becomes unrealistic for such colloids as Janus particles (JP),\cite{WaltherMuller2008,WaltherMuller2013} which have opened a new field of investigation with both fundamental and practical perspectives. Such JPs can be used for optical nanoprobes,\cite{Granick2008} E-paper display technology,\cite{Yin2011} or cargo transport.\cite{Golestanian2007,Liebchen2015} Their suspensions demonstrate rich phase behavior ranging from cross-linked gels up to ferroelectric crystals.\cite{Dempster2016} In the case of JPs, other factors like surface charge (or zeta potential) heterogeneity and anisotropy, come into play, so they should become a very important consideration in electrophoresis.

\begin{figure}[t]
\centering
  \includegraphics[scale=0.14, clip=true,trim= 0.0 13.0cm 0.0 8.7cm]{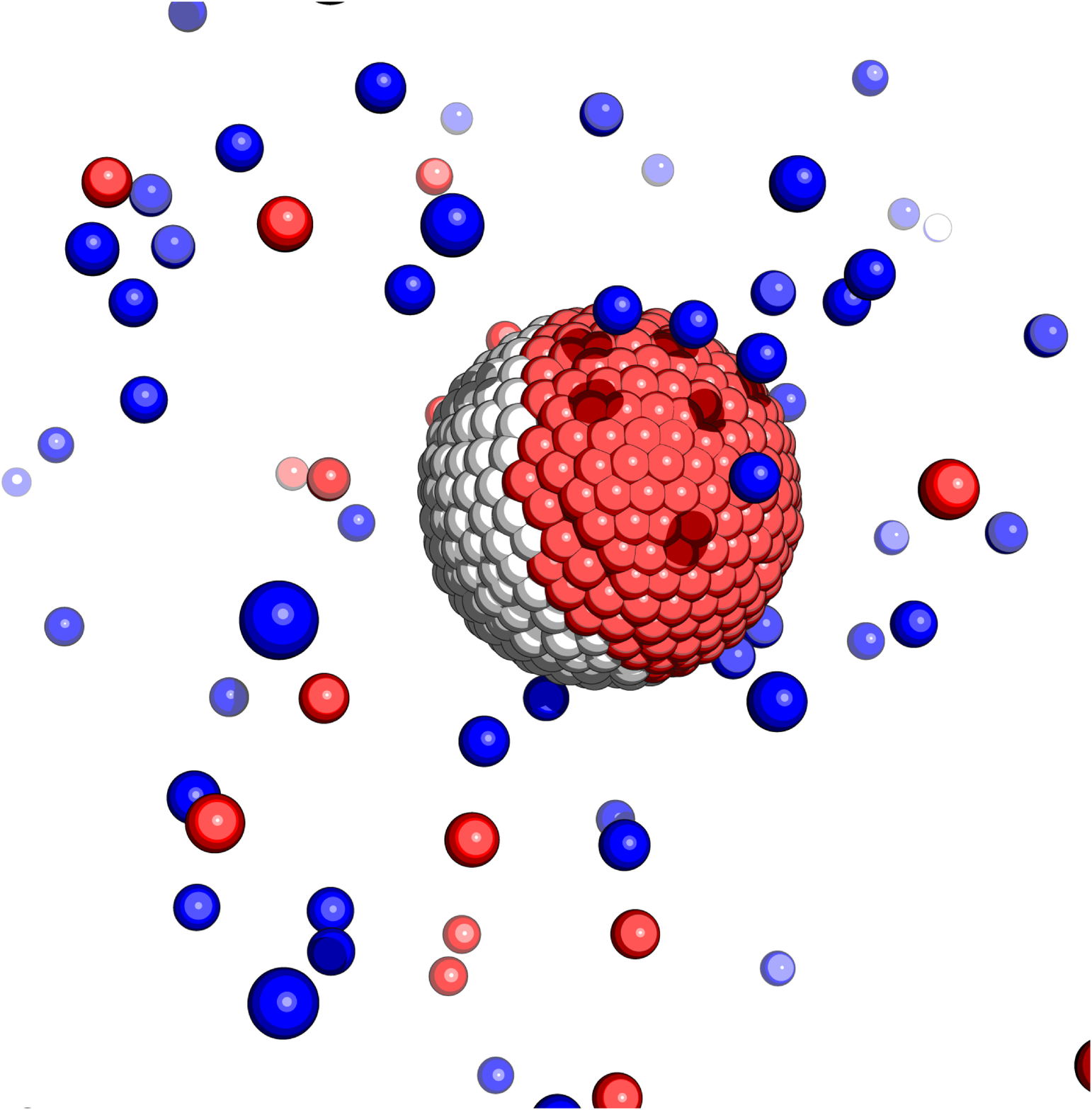}
  \caption{A snapshot of the simulation system: The `raspberry' colloid particle is surrounded by cations (red) and anions (blue). The uncharged sector is shown by white. The solvent grid is not shown.}
  \label{fig:raspberry}
\end{figure}

The body of theoretical and experimental work investigating electrophoretic properties of this class of particles is much less than that for uniform objects, and quantitative understanding of electrophoresis of JPs is still challenging, despite some recent advances. Previous theoretical work\cite{Anderson1985} has shown that in the thin EDL limit the electrophoretic mobility of JPs is well represented by the Smoluchowski model,\cite{Smoluchowski1924} i.e. remains governed by the average zeta potential, while the dipole moment of JPs only affects the orientation of particles relative to the external field. A recent numerical study \cite{Hsu2012} performed under the assumptions of SEM has shown that at the same averaged surface potential, the mobility of JPs in a spherical cavity of arbitrary size is generally smaller than that of a uniformly charged particle and the difference becomes more pronounced with the increase in non-uniformity. Molecular dynamics simulations have also concluded that charge inhomogeneities could reduce the diffusion coefficient of nanoparticles in nanopores.\cite{Su2012} Nevertheless, the electrophoretic properties of JPs remain largely unexplored outside the range of applicability of SEM, when one has to consider the finite size of ions and particles' own thermal wobble that disturbs the preferred alignment to the external field. Furthermore, we would like to point out that SEM assumes the constant surface potential while often it is the charge density that is kept constant. A recent study \cite{Ohshima2010} has pointed it out for the case of uniformly charged particles, but we  are  unaware of any previous work that has applied a constant charge condition for JPs.
%


%

\begin{figure}[t]
	\centering
	\includegraphics[scale=0.25, clip=false,trim= 0.0 0.0cm 1.0 1.0cm]{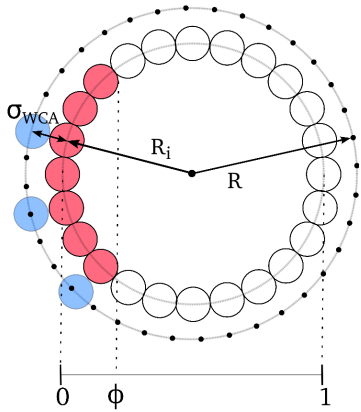}
    \caption{\label{fig:chi} Sketch of the model JP with  the radius of inner shell $R_{i}$,
      the effective cutoff radius for the WCA potential $\sigma_{WCA}$, and the radius of the outer shell $R=R_{i}+\sigma_{WCA}$ (black dots) serving as the effective hydrodynamic radius. The charged area of the fraction $\phi$ is shown by filled red circles. Open circles indicate uncharged region. Blue circles mark the closest distance at which ions do not experience any WCA repulsion from the colloid beads.}
    \label{fig:phi-def}
\end{figure}

In this paper, we use a hybrid Molecular Dynamics - Lattice Boltzmann simulation and a SEM-based mean-field approach to study the electrophoresis of a single JP.  We are interested in effects arising from variations of surface charge which occur both on the scale of particle radius and over distances comparable to the Debye length, so that we focus on intermediate values of $\kappa R = O(1)$, where quantitative understanding of electrophoresis remains especially challenging. Our results show that in at low total charges  JPs can be characterized by area-averaged surface charge, but at higher charges and $\kappa R$ their electrophoretic mobility is reduced, being strongly affected by non-uniformity and anisotropy of surface charge.

Our paper is organized as follows. In Section \ref{sec:method} we describe our methods and model of JPs. In section \ref{sec:result}, we present our data on electrophoretic mobility and rotational dynamics for JPs of different total charges and in various screening regimes. Our conclusions are summarized in Section \ref{sec:conc}.

\section{Simulation and numerical methods}
\label{sec:method}

For simulation of the dynamics of charged JPs, we use the hybrid Lattice-Boltzmann (LB) - Molecular dynamics (MD) method combined with the primitive model of the electrolyte.\cite{Duenweg1999} All MD-LB simulations are performed using ESPResSo.\cite{limbach06a,espresso2} We model all the charged species and the particle surface elements explicitly as MD beads, while the medium is modelled as a viscous fluid of mass density $\rho$ and dynamic viscosity $\eta$ at the level of the LB method. It is treated as a dielectric continuum characterized by the Bjerrum length $l_B$.

In our model, the MD beads -- small ions in solution and surface beads of the colloid -- interact via the Weeks-Chandler-Anderson (WCA) potential
\begin{equation}
	\label{eq:LJ}
	U_{WCA}(r) = \begin{cases} 4\epsilon_{ij}\left(\left(\frac{\sigma_{ij}}{r}\right)^{12} - \left(\frac{\sigma_{ij}}{r}\right)^6 + \frac{1}{4}\right), & r <2^{1/6}\sigma_{ij}, \\ 0, &  r \geq 2^{1/6}\sigma_{ij} \end{cases}
\end{equation}
and the Coulomb potential
\begin{equation}
	\label{eq:coulomb}
	U_C(r)= l_B k_B T \frac{z_i z_j}{r}.
\end{equation}
Here, $l_B$ is the Bjerrum length, $l_B = e^2/ (4 \pi \epsilon_0 \epsilon k_B T)$, and $z_i,z_j$ are ion valencies; $\epsilon_0$ and $\epsilon$ being the dielectric permittivity of vacuum and dielectric constant of water, respectively. The bead size, $\sigma_{ij}$, sets the unit length in our simulations and charachterstic energy scale is $\epsilon_{ij}=1$. $r$ is the distance between two MD beads. To facilitate a comparison to the mean-field theory we choose $\sigma_{ij}=2^{-1/6}\simeq 0.89\sigma$ so that $\sigma_{WCA}=1.0$ (Figure \ref{fig:phi-def}).

\indent
For the JP itself, we use the modified `raspberry' model of a spherical colloid particle.\cite{Lobaskin2004,Lobaskin2007,Lobaskin2008,Raspberry1,Raspberry2} The `raspberry' (Figure~\ref{fig:raspberry}) is a construct made of a single central MD particle with both translational and rotational degrees of freedom and two spherical shells around it made of `virtual' MD beads, whose positions are derived relatively to the central particle and not from the integration of their equations of motion. The inner shell has a radius of $R_{i} = R - \sigma_{WCA}\mathbf{ = 3.0\sigma}$ and holds the charged beads. The outer shell's radius is simply $R=4.0\sigma$: it does not interact with other MD particles either via WCA or Coulomb potentials but only with the LB fluid, and thus serves to define the colloid's hydrodynamic radius. Using two shells shifted against one another is advantageous for tuning both `electrostatic' and `hydrodynamic' radii to the same value which allows for more convenient comparison of our results to the mean-field theory predictions.
The coupling between the LB and MD subsystems is realized via dissipative interactions as introduced in Ref. \cite{Duenweg1999}. The viscous friction term, given by $\mathbf{F}_{f}=-\Gamma\left(\mathbf{u}_{s}-\mathbf{u}_{f}\right)$, where $\mathbf{u}_{s},\mathbf{u}_{f}$ are velocities of the solute beads and the solvent, respectively, acts on the solute particles -- microions and colloid surface beads. An opposite force is applied to the solvent to ensure momentum conservation, and Gaussian white noise $\mathbf{F_{R}}$ with zero mean is added that satisfies the fluctuation-dissipation theorem through $\left<F_{\alpha}\left(t\right)F_{\beta}\left(t'\right)\right>=2\delta\left(t-t'\right)2\delta_{\alpha\beta}k_BT~\Gamma$. This coupling mechanism also works as a thermostat, keeping the temperatures of MD particles and the LB fluid the same. We choose the simulation units as ${k_B T/\epsilon_{ij} = 1}$, $\rho={1.0\sigma^{-3}}$, ${\eta=3.0 \sqrt{m\epsilon_{ij}}\sigma^{-2}}$, $l_B=1.0{\sigma}$ and LB lattice spacing $a=1\sigma$. In comparison to the original `raspberry' model,\cite{Lobaskin2004,Lobaskin2004a} we here introduce two different friction coefficients for the microions and the colloid surface beads: for the surface beads $\Gamma$ is set to $20$, at which point the dependence of hydrodynamic radius on $\Gamma$ is saturated enough to emulate no-slip boundary condition at the hydrophilic surface\cite{Raspberry1}; while for microions $\Gamma$ is set to $2$ thus ensuring that the ionic atmosphere is fairly mobile compared to the colloid. The resulting reduced diffusion constant for microions was $6\pi\eta l_B D /k_B T \simeq 14$ which has been calculated from system parameters following Ahlrichs and D\"{u}nweg,\cite{Duenweg1999} and verified in a simulation via diffusivity measurements.

\begin{figure}[t]
	\centering
	\includegraphics[scale=0.48, clip=true, trim=0.7cm 1.2cm 0.8cm 2.5cm]{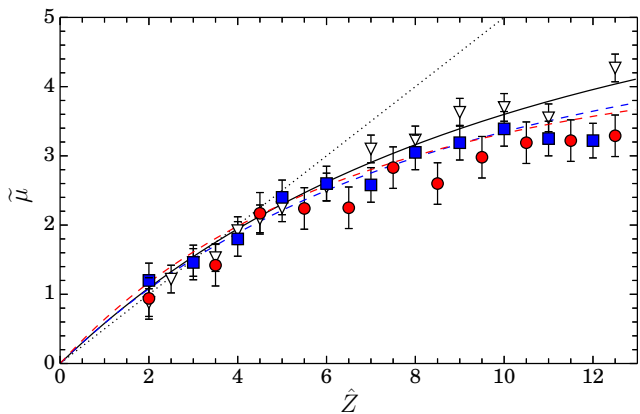}
	\caption{\label{fig:mob-charge}
 	Electrophoretic mobility of Janus particles with $\phi =1.00$ (open triangles), $0.50$ (filled squares), and $0.25$ (filled circles) as a function of net charge at $\kappa R = 1.0$. Solid curve shows the predictions of SEM, dashed curves represent numerical mean-field calculations for JPs, dotted line shows the H\"{u}ckel limit solution. The color of the curves matches the color of the symbols for the corresponding charge distribution.}
\end{figure}

The system including the particle, electrolyte, and fluid was modeled in 3D periodic boundary conditions in a cubic box with $L=40.0\sigma$, giving the $\mathbf{R/L=0.1}$ and the colloid volume fraction of 0.41\%. The number of monovalent ions $N$ in the simulation box was set by the number of background salt ions and counterions to the colloid, $N = 2 C_{\Sigma} L^3 + |Q/e|$, where $Q$ is the colloid charge so that the system was overall electroneutral. The electrostatic interactions were evaluated using P3M implementation of the Ewald summation technique.\cite{Ballenegger2008} We describe the ionic strength and the screening conditions in the suspension by $\kappa R$, where $\kappa = (4 \pi l_B N/L^3)^{1/2}$, with the total number of the (monovalent) ions in the simulation box $N$, and we vary $\kappa R$ from $0.5$ to $3.0$ through the concentration of the salt ions, which is typically $0.001-0.025 \sigma^{-3}$. The external field $E$ was modeled by a uniform force acting on each charged MD bead, and in all simulations we use $E=0.2 k_BT/\sigma e$. This field strength belongs to the  linear response regime for our systems, which is confirmed by the linear dependence of the velocity on the field strength, i.e. constant mobility. At the same time, the field is sufficiently large to give a noticeable particle velocity and to facilitate the mobility measurements. The ionic cloud at this field value is not significantly perturbed, while the external field $E$ is less than the potential drop over the electrostatic diffuse layer $\kappa\zeta$. The chosen field strength is also suitable to study the interplay between the JPs' thermal wobble and electrostatic torque, which will be described in details in the following sections.

\begin{figure}[t]
	\centering
		\includegraphics[scale=0.48, clip=true, trim=0.7cm 1.4cm 0.8cm 2.5cm]{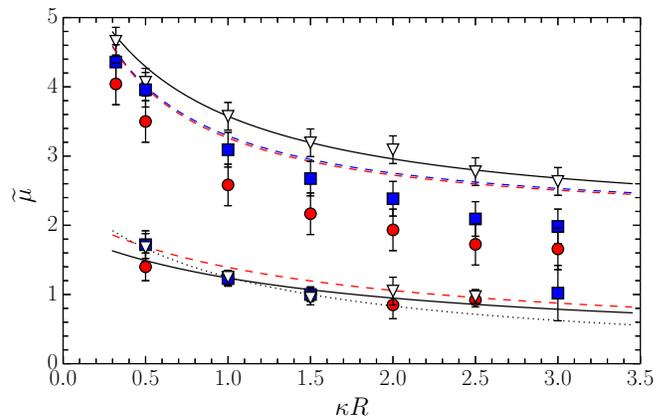}
   \caption{\label{fig:mob-kr-z9.6}
   		Electrophoretic mobility of particles as a function of $\kappa R$ at high,  $\hat{Z}=9.6$ (top datasets) and low $\hat{Z}=2.5$ (bottom datasets) charge regimes. Symbols show simulation data obtained at $\phi=1.00$ (triangles), $0.50$ (squares), and $0.25$ (circles). Solid curve shows predictions of SEM, dashed curves represent numerical results for JPs, dotted line shows the H\"{u}ckel limiting solution for uniformly charged colloid. The color of the curves matches the color of the symbols for the corresponding charge distribution.}
\end{figure}

\begin{figure*}[t]
	\centering
	\includegraphics[scale=0.85, clip=true, trim=0.2cm 4.cm 0.0cm 4.8cm]		{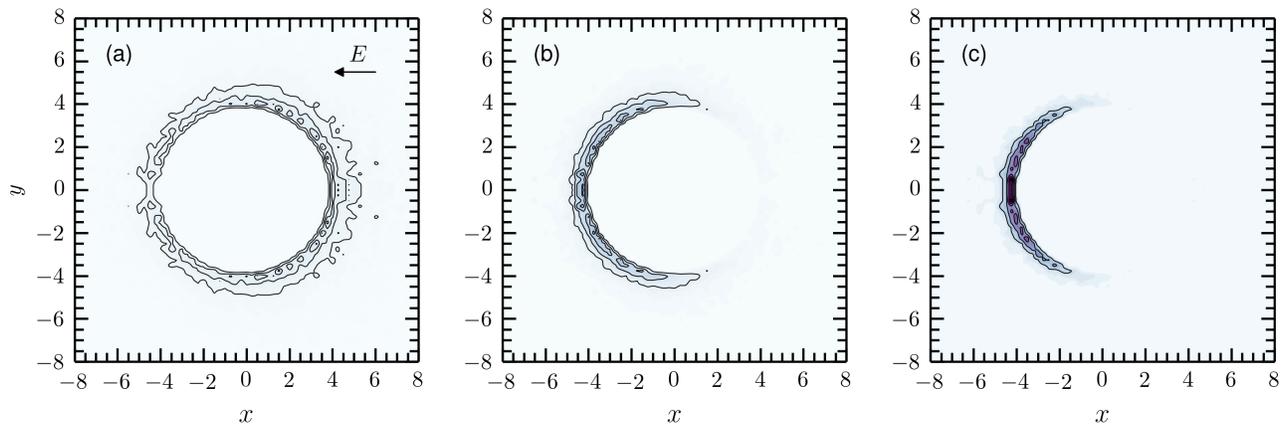}
	\caption{\label{fig:2D-ions}
 	2D radial axisymmetric distribution maps of counter-ions around particles of $\phi=1$ (a), $0.50$ (b), $0.25$ (c) calculated at $\hat{Z}=9.6$ and $\kappa R=1.0$.  }
\end{figure*}

The area fraction of the charged surface, $\phi$, (Figure~\ref{fig:chi}) has been varied  from $0.25$ to $1$, where the latter corresponds to a uniformly charged particle, by keeping the net charge of the colloid $Q$ constant. The charged patch was always a spherical segment of a given height $2\phi R$ (Figure~\ref{fig:chi}), so all the JPs had axial symmetry of charge distribution. We should stress that  all the JPs had the same net charge and the same average charge density $\langle q\rangle = Q/4 \pi R^2$, but different local charge densities $q = \langle q\rangle/\phi$ of their `patches'. Further in the text, we use the scaled charge $\hat{Z}\equiv Q\frac{l_B}{e R}$, which we vary from $2.0$ to $12.5$ by changing $Q$ via the charges of surface beads.

We also employ a direct numerical solution of the electrophoretic problem that we describe in Appendix \ref{sec:solver}. We the reader for a more in-depth look on this system in the original publication\cite{Schmitz2012}. We use such an approach for precise control over the charge distribution around the colloid, and accordingly we implement it to construct JPs in exactly the same manner as in MD-LB model, i.e. by distributing fixed charge over fraction of the surface, $\phi$.

\section{Results and discussion}
\label{sec:result}

\subsection{Electrophoretic mobility}
\label{sec:mu}

We first investigated the effect of a charge heterogeneity on the electrophoretic mobility. Figure~\ref{fig:mob-charge} shows the simulation results for electrophoretic  mobility as a function of $\hat{Z}$ obtained for $\kappa R = 1.0$ at different fractions of charged area, $\phi$. Also included are numerical mean-field results and predictions of the H\"{u}ckel theory $\widetilde{\mu}=\hat{Z}/\left(1+\kappa R\right)$.\cite{Huckel1924} We see that when the particle is uniformly charged ($\phi = 1$), at relatively small charges the mobility is nearly equal to, while at high charges is smaller than that predicted in the H\"{u}ckel limit. This confirms that the electrophoretic mobility of a uniformly charged particle is proportional to its charge only in the weak charge regime.\cite{OBrienWhite1978,Ohshima2010} Note that the simulation results for a uniformly charged particle are in excellent agreement with predictions of the SEM, which demonstrates the predictive power of our simulation model. Another result emerging from Figure~\ref{fig:mob-charge} is that in the low $\hat{Z}$ regime simulation data at $\phi=0.50$ (a `balanced' charge distribution) and $\phi=0.25$ (highly concentrated charge on a relatively small surface patch) practically coincide with those obtained for the uniformly charged particle. This indicates that the electrophoretic mobility in this regime is fully determined by the area-averaged charge (or zeta potential) in agreement with the SEM for uniformly charged particles, as it is commonly assumed. In the high charge regime, the electrophoretic mobility of JPs is smaller than that of a uniformly charged particle, so that the SEM for the uniformly charged colloid significantly overestimates simulation results. This means that the mobility is no longer determined by the area-averaged charge alone. Our observation -- that the electrophoretic mobility of JPs decreases at high charges -- is likely related to the non-linear relation between $\widetilde{\mu}$ and $\hat{Z}$. Since all the JPs we study bear the same net charge, their local charge densities vary significantly. Hence in the regime of high charges the mobilities of JPs divert from the linear relation even more than it might be expected for the uniformly charged particles.

The decrease in the electrophoretic mobility of JPs is also captured by our SEM calculations. We remark, however, that our numerical solutions show practically no difference in electrophoretic mobilities for particles of $\phi=0.50$ and $0.25$, but the deviations of simulation data from the SEM are getting larger when $\phi=0.25$. We see that the SEM and primitive model simulation results for JPs practically coincide at $\phi=0.5$ (except for the highest tested $\hat{Z}$), but at $\phi=0.25$ the simulation data deviate from the numerical solution towards smaller mobilities. We can speculate that numerical solutions deviate from the MD-LB simulations (at high enough charges or strong enough screening) because of the limited resolution of the lattice that the mean-field solver we use provides. Indeed, when $\kappa R$ grows, the characteristic thickness of the electrostatic diffuse layer decreases, and with high charges the potential grows too rapidly in the close vicinity to the charged surface. It is well-known\cite{borukhov2000} that Poisson-Boltzmann equation often may not grasp this rapid increase properly, thus decreasing the accuracy of the results. The fact the our MD-LB results deviate even more from the numerical solution in case of $\phi=0.25$ is consistent with this suggestion, since the local density is fairly high in this case to be precisely resolved by lattice-based solver.

It is instructive now to focus on the role of $\kappa R$. Figure~\ref{fig:mob-kr-z9.6} shows numerical and simulation results obtained at low, $\hat{Z}=2.5$, and high, $\hat{Z}=9.6$, values of surface charge. A general conclusion from this plot is that in this range of parameters the electrophoretic mobility decays with $\kappa R$, but the influence of charge non-uniformity is different for low and high net surface charge.
In the case of small charges, the mobility of JPs does not significantly differ from that of uniformly charged colloids, and the simulation data are in agreement with mean-field theory results. One can therefore conclude that a simple H\"{u}ckel model can safely be used to analyze the mobility data in the studied range of $\kappa R$. In the high charge regime, we see that the simulation data for uniformly charged particle are well fitted by the SEM, and are well below the H\"{u}ckel solution. For JPs the numerical solution predicts slightly lower electrophoretic mobility, actually the same for $\phi=0.50$ and $0.25$. The simulation data deviate from these mean-field solutions towards the smaller mobility values, especially at larger $\kappa R$. We also note that the discrepancy is larger for JPs of $\phi=0.25$.

\begin{figure}[t]
	\centering

 	\includegraphics[scale=0.48, clip=true, trim=0.7cm 1.2cm 0.0cm 2.5cm]{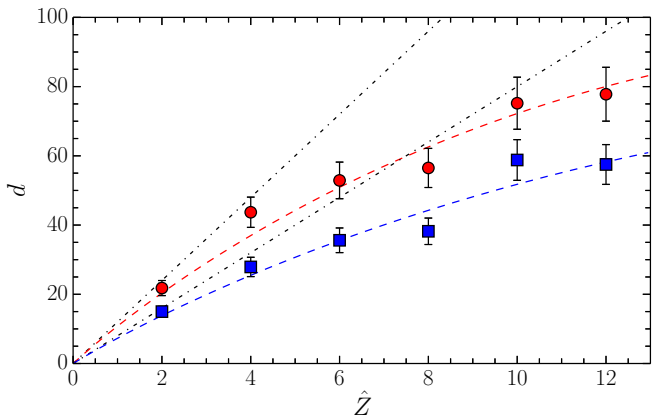}

   \caption{\label{fig:dipole}
   		Dipole moment of JPs with $\phi=0.25$ (circles) and $0.50$ (squares) measured from their rotational `wobble' at $\kappa R = 1$. Dash-dotted lines show dipole moment of unscreened particles. Dashed curves are drawn only to guide the eye.}
\end{figure}



\subsection{Orientation and dipole moment}

We present  the counterion density maps near particles of different $\phi$ in Figure~\ref{fig:2D-ions}. This plot demonstrates that JP dipole moments are oriented along the external field. Note that our density maps account for the JPs own wobble, so that the thermal motion perturbs the preferred orientation of the particle, which means that counter-ion cloud also experiences orientational fluctuations. We also remark that the accumulation of counterions near JPs is stronger compared to a uniformly charged colloid, and increases with a decrease in $\phi$. This obviously reflects the fact that at the same $\hat{Z}$ the local charge density of the charged area is higher at smaller $\phi$. These results indicate a non-negligible dipole moment of JPs, which should correlate with the decrement of electrophoretic mobility values, as both are caused by the charge screening. Motivated by these observations below we studied the orientational torque of JPs in the electric field and its relation to the mobility.

\begin{figure}[t]
	\centering
	\includegraphics[scale=0.48, clip=true, trim=1.2cm 2.0cm 0.0 2.6cm]{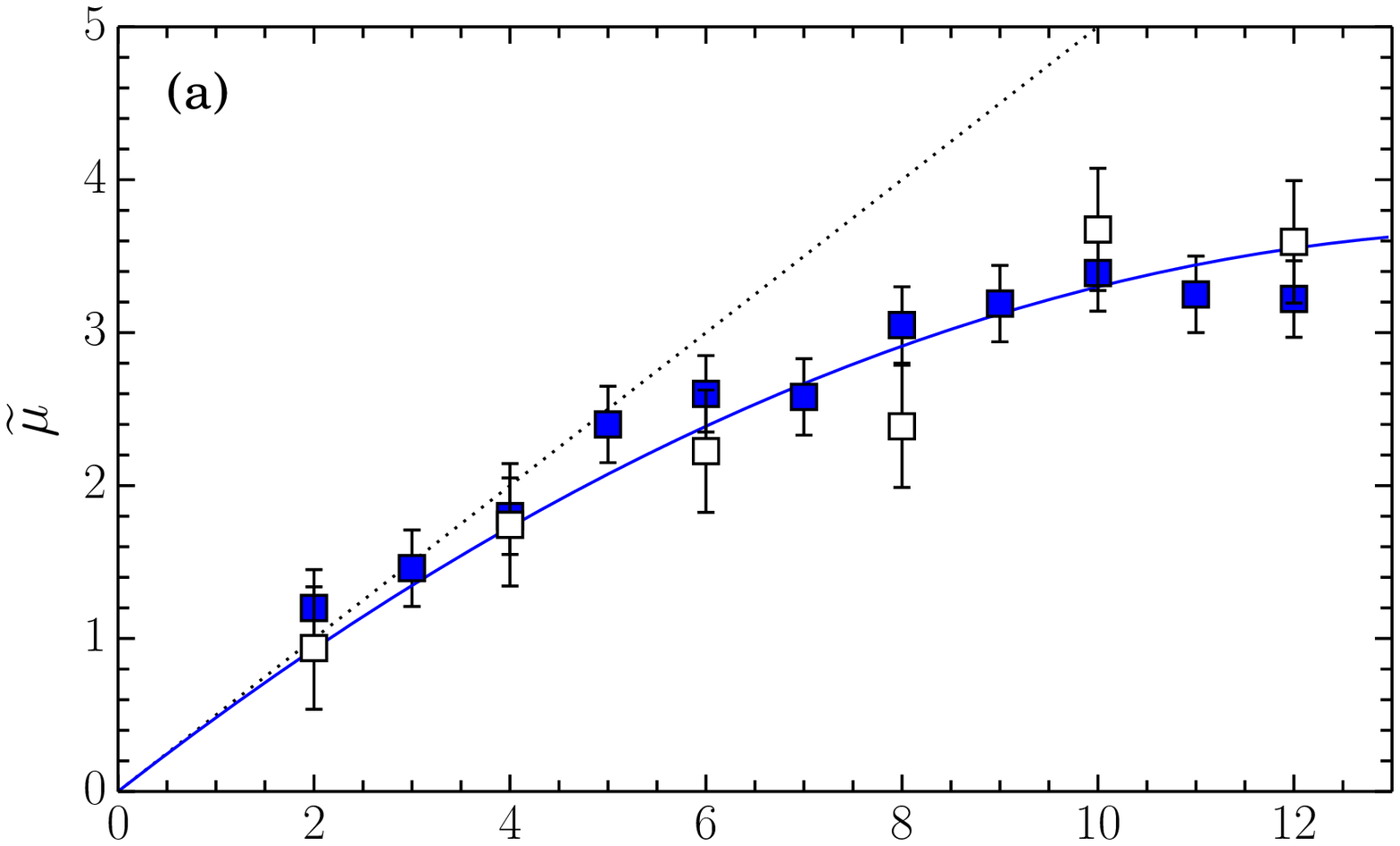}
	\includegraphics[scale=0.48, clip=true, trim=1.2cm 1.6cm 0.0cm 2.2cm]{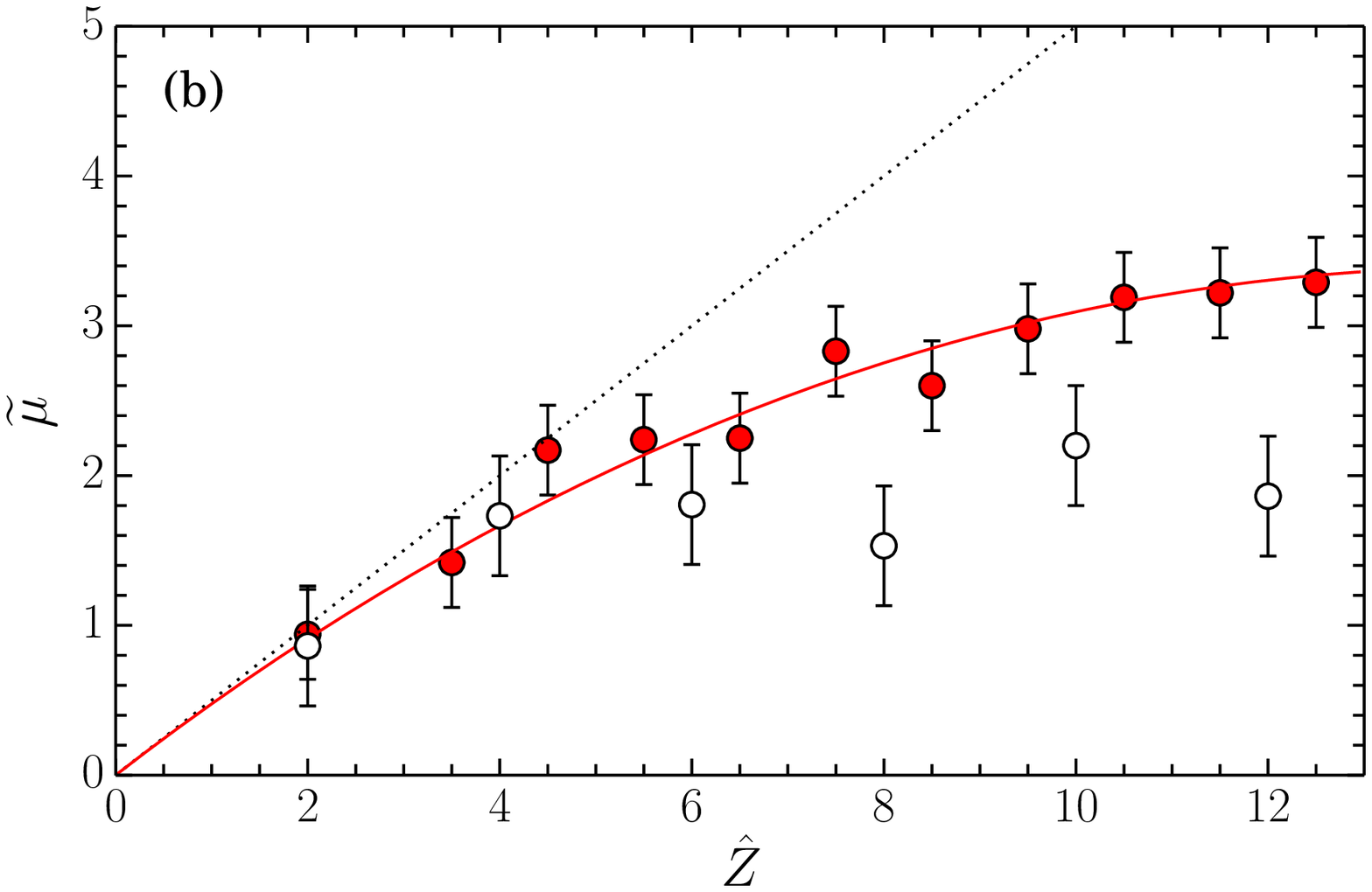}
	
   \caption{\label{fig:effect}
   		Electrophoretic mobility at $\kappa R = 1$ calculated from dipole moments of JPs with $\phi=0.50$ (a) and $0.25$ (b) as a function of net charge (empty symbols). The same data as in Figure~\ref{fig:mob-charge} is shown by blue squares and red circles respectively. Dotted lines plot the H\"{u}ckel limit solution \eqref{eq:DH-mob}, solid curves are drawn only to guide the eye.}
\end{figure}

The JP's rotational dynamics can be characterized by the mean-squared angular displacement $\left<\alpha^2 \left(\tau \right) \right>$ of the particle's dipole moment within time $\tau$. Our analytical solution is described in Appendix~\ref{sec:orient}, and can be presented as
\begin{equation}
	\label{eq:langevin_main}
	\sum_{n=1}^{\infty} \left(-1\right)^{n+1} \frac{\left<\alpha^{2n}\right>}{\left(2n-1\right)!}_{\tau\rightarrow\infty} \equiv \left<\alpha\right>_{\Sigma} = \frac{k_BT}{Ed},
\end{equation}
where $d$ is the dipole moment, which can be deduced from  $\left<\alpha\right>_{\Sigma}$, calculated by using the simulation data on $\alpha$ as a function of time~$\tau$.

In Figure \ref{fig:dipole}, we plot the dipole moment of two JPs as a function of their net charge $\hat{Z}$. We note that at low $\hat{Z}$ the measured $d$ coincides with the dipole moment of the unscreened particle, $d_0$, which can be defined as $d_0 = \left(1-\phi\right)R^2 \hat{Z} e /l_B$. However, at high  $\hat{Z}$ it is considerably smaller, which immediately suggests that counterions contribute to its effective value, likewise they contribute to the effective zeta potential and to the decrement of electrophoretic mobility.

To test this assumption, we attempted to predict the values of electrophoretic mobility from the measured dipole moment. We introduce an effective charge as $\hat{Z}_{\text{eff}}=\hat{Z} d/d_0$, and then use it to compute the mobility using the H\"{u}ckel limiting law for low $\hat{Z}$ and $\kappa R$
\begin{equation}
\label{eq:DH-mob}
	\tilde{\mu} = \frac{\hat{Z}}{1+\kappa R}.
\end{equation}

\noindent
In Figure~\ref{fig:effect}, we plot the  mobilities calculated at $\phi=0.50$ and $0.25$ and compare them with data from Figure~\ref{fig:mob-charge}. We see that for $\phi=0.50$ (Figure~\ref{fig:effect}(a)) the two sets of data agree with each other quite well, so that the measurements of dipole moments can be used to evaluate the decrement of electrophoretic mobility of JPs. However,  for $\phi=0.25$ (Figure~\ref{fig:effect}(b)) the effective charge approach underestimates the mobility at high $\hat{Z}$. Since even at the highest local charge density the hydrodynamic radius (calculated with Eq.(\ref{eq:lambda}) of Appendix~\ref{sec:orient}) remains the same within a statistical error, such a discrepancy cannot be related to counterion condensation. So, it is likely that the concept of single effective charge becomes unsuitable when the surface charge anisotropy is getting very large.

\section{Conclusions}
\label{sec:conc}

We have studied the electrophoretic mobility of JPs  and have shown that it depends both on their net charge and charge distribution. Namely, less homogeneous charge distributions generally lead to lower mobilities, which is consistent with previous observations made for different systems.\cite{anderson.jl:1989,ajdari2002,Vinogradova2011} The decrease in mobility as compared to that of uniformly charged particles is negligibly small at low particle net charges and small $\kappa R$. In this case, the electrophoretic mobility can be related to the area-averaged charge (or zeta potential) thought the SEM, as it is commonly assumed. The deviations from the SEM are becoming pronounced when the net charge and $\kappa R$ increase. Where the mobility is significantly affected by charge heterogeneity, the mean-field predictions for JPs overestimate the mobility, and should be used with care. Reversely, the zeta potential or surface charge extracted from the mobility data in the regime $\kappa R \approx 1$ for nanoparticles and molecules with non-uniform surface charge distribution with SEM are expected to underestimate the particle net charge. We have also shown that JPs' dipole moments align to electric field, and that their orientation and dipole moment are strongly correlated with the electrophoretic mobility and can be used for predicting the mobility decrease.

\section{Acknowledgements}

This research was supported by the Russian
Academy of Sciences (priority programme `Assembly and Investigation of Macromolecular Structures of New Generations').
The simulations were carried out using computational resources at the Moscow State University (`Lomonosov' and `Chebyshev').
We have benefited from discussions with Jiajia Zhou, Salim Maduar and Alexander Dubov. We thank Roman Schmitz and Burkhard D\"{u}nweg for access to their SEM solver.

\appendix
\section{\label{sec:solver}Numerical solution of mean-field equations}

The algorithm we apply is a solver for the following Poisson-Boltzmann equation and a coupled set of Nernst-Plank and Stokes equations:

\begin{equation}
\label{eq:SEM1}
0 = \nabla^2 \psi + \frac{1}{\epsilon}e \sum_i z_i c_i,
\end{equation}

\begin{equation}
\label{eq:SEM2}
0 = \nabla \cdot \left(D_i\nabla c_i + \frac{D_i}{k_B T}e z_i \left(\nabla \psi\right)c_i - \mathbf{v}c_i \right),
\end{equation}

\begin{equation}
\label{eq:SEM3}
0 = -\nabla p + \eta \nabla^2 \mathbf{v} - e\left(\nabla \psi \right) \sum_i z_i c_i,
\end{equation}

\begin{equation}
\label{eq:SEM4}
0 = \nabla \cdot \mathbf{v},
\end{equation}

\noindent
where $z_i$, $c_i$ and $D_i$ are valencies, concentrations and diffusion constants of charged species $i$, $\psi$ is the electrostatic potential, $p$ is the pressure and $\mathbf{v}$ is the velocity of liquid in the fixed colloid's reference frame. In this method\cite{Roman2011}, instead of solving a set of partial differential equations for electrostatics, the problem is reformulated in terms of electric field rather than potential to obtain a free energy in the form of the following functional

\begin{equation}
\label{eq:free-energy1}	
	F=\int_V f dV \\
\end{equation}
\begin{align*}
\label{eq:free-energy2}	
	f = \frac{1}{2}\mathbf{E}^2 + &\sum_i c_i \ln c_i  - \psi \left(\nabla \cdot \mathbf{E} - \sum_i c_i \right) \\
	&-\sum_i \mu_i \left(c_i - \frac{N_i}{V}\right)
\end{align*}		

\noindent
where are total numbers of charged species $i$, and $V$ is the system volume. The minimum of this functional corresponds to the solution of a related Poisson-Boltzmann equation. Further discretization allows one to implement a version of this solver with charged species serving both as ions in solution and the surface charged beads of a colloid, in a sense much like the MD implementation, albeit a lattice one. In order to minimize the free energy functional the discrete charges are moved around the lattice.The solution is used as an input to the set of linearised Nernst-Plank and Stokes equations, and this procedure is repeated iteratively until the solution for the fluid velocity in the frame of the colloid converges.

\section{\label{sec:orient}Orientational dynamics of a Janus particle}

For the case of a constant and uniform electric field $\mathbf{E}$, a dipole orientation satisfies the Boltzmann distribution
\begin{equation}
	\label{eq:angle_dist_Boltzmann}
W = A e^{E d \cos \alpha/k_B T},
\end{equation}
where $A = 4 \pi \frac{k_BT}{ E d} \sinh{ \frac{E d}{k_B T}}$ is a normalisation constant. The average value of the dipole moment component in the direction of the field can be then calculated as
\begin{align}
	\label{eq:av_angle}
d   \langle \cos \alpha \rangle  &= \int_0^{2 \pi} \int_0^\pi W(\alpha) d \cos\alpha \sin\alpha d\alpha d\varphi \nonumber \\
 & = d  \left ( \coth{\frac{E d}{k_B T }} - \frac{k_B T}{E d} \right ).
\end{align}
In strong fields such that $\frac{E d}{k_BT} \gg 1$, the first term in the bracket turns unity and we have
\begin{equation}
	\label{eq:av_angle_as}
d  \langle \cos \alpha \rangle_{E \to \infty} = d \cdot \left ( 1 - \frac{k_B T}{E d} \right ),
\end{equation}
whence, since $\alpha$ is small, we find
\begin{equation}
	\label{eq:av_angle_alpha}
\langle \alpha^2 \rangle  =  \frac{k_B T}{E d}.
\end{equation}

We can write a 1D Langevin equation for the colloid orientation angle with respect to the field as
\begin{equation}
	\label{eq:langevin_base}
	I\ddot{\alpha} = -Ed\sin\alpha - \zeta_R \dot{\alpha} + T_{rand}(t).
\end{equation}
Here $I$ stands for the colloid's moment of inertia, $\zeta_R=8\pi\eta R^3$ and dot and double dot denote the first and second time derivatives, respectively. Multiplying \eqref{eq:langevin_base} by $\alpha$, expanding $\sin\left(\alpha\right)$ in a series and also using that $d\left(\alpha\dot{\alpha}\right)/dt\equiv\dot{\alpha}^2+\alpha\ddot{\alpha}$, we find
\begin{equation}
	\label{eq:langevin_2}
	I\frac{d}{dt}\left(\alpha\dot{\alpha}\right)-I\dot{\alpha}^2 = -Ed\sum_{n=1}^{\infty} \left(-1\right)^{n+1} \frac{\alpha^{2n}}{\left(2n\right)!} - \zeta_R\dot{\alpha}\alpha + T_{rand}\alpha,
\end{equation}
which after taking the ensemble average of both parts becomes
\begin{equation}
	\label{eq:langevin_3}
	\frac{I}{2}\frac{d^2}{dt^2}\left<\alpha^2\right> + \frac{\zeta_R}{2}\frac{d}{dt}\left<\alpha^2\right> + Ed\sum_{n=1}^{\infty} \left(-1\right)^{n+1} \frac{\left<\alpha^{2n}\right>}{\left(2n\right)!} - k_BT = 0.
\end{equation}
Here we have used the equipartition theorem, $T_{rand}$ symmetry and the fact that $\left<\alpha\dot{\alpha}\right>=\frac{1}{2}\frac{d}{dt}\left<\alpha^2\right>$.
In the limit $\alpha\rightarrow 0$, we can further simplify Eq.\eqref{eq:langevin_3} by omitting higher order terms and using $\mathds{A}=\left<\alpha^2\right>-k_BT/Ed$:
\begin{equation}
	\label{eq:langevin_4}
	\ddot{\mathds{A}} + \frac{\zeta_R}{I}\dot{\mathds{A}} + \frac{2Ed}{I}\mathds{A} = 0.
\end{equation}
This is a well-known differential equation for a damped oscillator\cite{LandafchitzVol1}, in our case -- over-damped as the damping parameter $\zeta_R/\sqrt{8IEd}$ is always larger than $1$. Therefore, we can write the solution in the form of $\mathds{A}=Ae^{-\lambda t}$ where $\lambda$ is the smaller root of an auxiliary quadratic equation (we drop one term with large $\lambda$ as it decays too fast and is negligible). Finally,  we use the condition $\left<\alpha^2\right>_{t=0}=0$ to get
\begin{equation}
	\label{eq:langevin_5}
	\left<\alpha^2\right>\left(\tau\right) = \frac{k_BT}{Ed}\left(1-e^{-\lambda \tau}\right).
\end{equation}
This equation describes the evolution of the JP's orientation relative to the external field and at $t\>>1/\lambda$ the result agrees with Eq.(\eqref{eq:av_angle_alpha}) derived from the statistical viewpoint. Substituting the solution for $\left<\alpha^2\right>$ in equation \eqref{eq:langevin_3} we see that both derivatives vanish when $t\rightarrow\infty$ and thus we can find the full solution in the equilibrium distribution without the limitation $\alpha \rightarrow 0$:
\begin{equation}
	\label{eq:langevin_6}
	\sum_{n=1}^{\infty} \left(-1\right)^{n+1} \frac{\left<\alpha^{2n}\right>}{\left(2n\right)!}_{\tau\rightarrow\infty} \equiv \left<\alpha\right>_{\Sigma} = \frac{k_BT}{Ed}.
\end{equation}
Since $\lambda$ is a function of both $\zeta_R$ and $I$ we can rewrite the solution of auxiliary equation in terms of $R$ and omit the $\lambda^2$ term
\begin{equation}
	\label{eq:lambda}
	R \simeq ({\frac{Ed}{4\pi\eta\lambda}})^{1/3}.
\end{equation}
Thus, the same $\langle \alpha \rangle$ data allow one to calculate the hydrodynamic radius of the particle.




\bibliographystyle{jcp} 

\end{document}